\documentclass[aip,reprint]{revtex4-1}
\usepackage{color}
\usepackage{graphicx}
\usepackage{dcolumn}
\usepackage{bm}
\usepackage{sidecap}
\draft 

\begin{document}

\title{Numerical modeling of active thermo-plasmonics experiments} 

\author{Giuseppe Emanuele Lio}
\email[]{E-mail: giuseppe.lio@unical.it}
\author{Giovanna Palermo}
\author{Antonio De Luca}
\affiliation{CNR-Nanotec, 87036 Arcavacata di Rende (CS), Italy}
\affiliation{Physics Department, University of Calabria, 87036 Arcavacata di Rende (CS), Italy}
\author{Roberto Caputo}
\affiliation{CNR-Nanotec, 87036 Arcavacata di Rende (CS), Italy}
\affiliation{Physics Department, University of Calabria, 87036 Arcavacata di Rende (CS), Italy}
\affiliation{Institute of Fundamental and Frontier Sciences, University of Electronic Science and Technology of China, Chengdu 610054}

\begin{abstract}
In this paper, we present a simple and robust numerical method able to predict, with high accuracy, the photo-thermal effects occurring for a gold nanoparticles arrangement under externally applied strain. The physical system is numerically implemented in the COMSOL Multiphysics simulation platform. The gold nanoparticles distributions are excited by linearly polarized light. By considering the system at rest and under the action of a mechanical stress, we analyze the extinction cross section, and we observe the production of heat at the nanoscale. The purpose of this work is to describe how sensitive the local temperature of the gold nanoparticles arrangement is to the formation of localized photo-thermal hot spots. 
\end{abstract}


\maketitle 

\section{Introduction}
In the last decade, the multi-disciplinary character of research in materials science is gradually increasing the degree of complexity of the systems under study. An analytical approach to the modeling of such systems is de facto leaving the field to a numerical counterpart. This trend in research is thus promoting the evolution of comprehensive numerical packages providing an extremely accurate description of physical processes.
As a consequence, the availability of such powerful tools are modifying the way research is performed, eventually suggesting new possibilities to optimize the scientific procedure itself.
An emerging direction in plasmonics is the development of plamonic heaters intended as systems where the heat is generated at the nanoscale, by light excitation of plasmonic sub-entitites, that can be dynamically controlled in intensity and direction of its flow. Here, we present the performance of a specific design of plasmonic heaters through a numerical characterization performed by the COMSOL Multiphysics FEM commercial platform.
As reported in previous works, it is possible to evaluate the photo-thermal effects related to the interaction of an electromagnetic wave with an arrangement of gold nanoparticles (GNPs) arrangements. The behavior of the considered system can be accurately predicted when the GNPs is modified because of the application of a mechanical tensile strain to the whole structure. \cite{palermo2018flexible, lio2019opto} 
By consecutively exploiting Structural Mechanics, Electromagnetic Waves and Heat Transfer modules of the COMSOL software allows calculating: i) the new GNPs displacements as a function of a specific percentage of strain; ii) the plasmonic response of the GNPs distribution for ``at rest'' and ``stretched'' conditions, in terms of extinction cross section ($\sigma_{ext}$); iii) the temperature variation $\Delta T=T-T_0$, with respect to the environment temperature ($T_0$), measured when the GNPs geometry is resonantly excited by a linearly polarized plane wave. Considered the fine control of the local temperature allowed by plasmonic heaters, their exploitation can widen the actual scenario of thermo-plasmonic application in cellular biology,  nano-medical diagnostics and therapy. In fact, most fascinating challenges could include but are not limited to the design of photo-thermal devices with a high localized photo-thermal efficiency, controllable through external macroscopic stimuli.
Although several theoretical and experimental studies\cite{baffou2010thermoplasmonics, govorov2014photogeneration, lio2019tensile}  are all aimed to go in this direction, the actual possibility to realize mechanically tunable plasmonic nano-heaters by exploiting plasmo-mechanics\cite{maurer2015beginnings,cataldi2014growing, gontier2017optical, marae2018dense} and active plasmonics \cite{macdonald2010active, schuller2010plasmonics, caputo2013active} is still an open challenge.
\section{Results and Discussion} 
In this section, we show how it is possible to reproduce an entire experimental procedure in a numerical way. We will start to approach the experiment by applying an external strain to modify the initial arrangement of the considered GNPs supported by an elastic substrate. The geometry at rest is a square of 25 NPs, characterized by a radius ($R$) of 20 nm and center-to-center interdistance  $r$ equal to 3$R$. The first step consists in the application of the mechanical strain, followed by the electromagnetic modeling. Finally, it is shown how the opto-mechanical effects are exploited to enhance the photo-thermal response of the structure for different amount of stretching. 
\subsection{Mechanical strain}
\begin{figure}[h]
\begin{center}
\includegraphics[width=1\columnwidth]{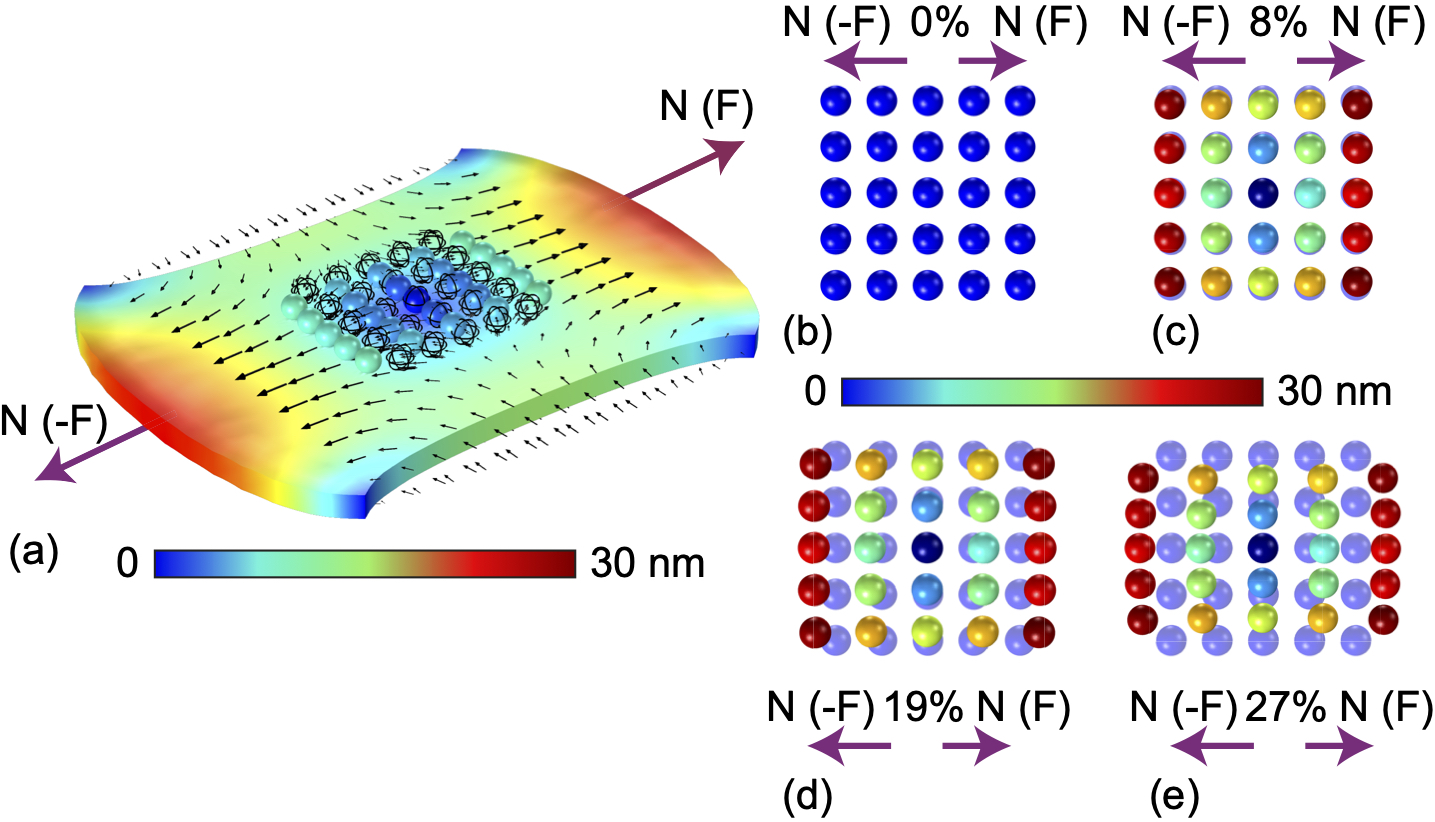}
  \caption{(a) Sketch of the entire system (GNPs/substrate) under stretching.  The edges where the loads assume a red coloration indicate the maximum displacement and are deformed in accordance with the strain direction. On the other hand, the other edges of the substrate appear green because of the compression undergone in the direction perpendicular to the applied strain. (b) The GNPs arrangement at the rest (b) and for the stretching percentage of (c) 8\%, (d) 19\% and (e) 27\%.  The blurred blue spheres in c-e represent the GNPs position at rest.}
\label{1}
\end{center}
\end{figure}

In a real experiment, it is possible to apply a strain to a sample in different ways, the more common is to use a tensile-strain machine. In our numerical experiments the external strain is applied as a boundary load at the edge of the elastic substrate. As in the reality, the GNPs arrangement is placed onto a transparent and elastic polymer matrix made of Polydimethylsiloxane (PDMS). The stretching appears when we apply a normal (\textbf{N}) tensile strains with positive (\textbf{F}) and negative forces (\textbf{-F}) along the $x$ direction.
It produces the effect shown in Figure \ref{1}a, where it is possible to evaluate the stretching of the substrate. In our case we are interested on the detailed displacement of each GNP. Figure \ref{1}b shows the GNPs arrangement at rest, while in Figure \ref{1}c-e are shown the displacements of the GNPs for stretching percentages of 8\%, 19\% and 27\%, respectively. 

\subsection{Electromagnetic analysis}
 When the new positions of the GNPs in the stretched geometries are calculated, we can use them to study what happens when a polarized electromagnetic wave impinges on the resulting different particles arrangements. To study the plasmonic response of the GNPs arrangements (at rest and stretched) we collect the absorption cross section $\sigma_{abs}$ and the scattering cross section $\sigma_{sca}$ to obtain the extinction cross section: $\sigma_{ext}=\sigma_{abs}+ \sigma_{sca}$.
In the methods section we report all details related to the strategy that we adopted to analyze $\sigma_{ext}$ and the materials involved in our model. 
We stress out that by modifying the GNPs arrangements through an externally applied strain, it is possible to change the wavelength corresponding to the plasmonic resonance $\lambda_{p0}$ of the initial system. 

\begin{figure}[h]
\begin{center}
\includegraphics[width=1\columnwidth]{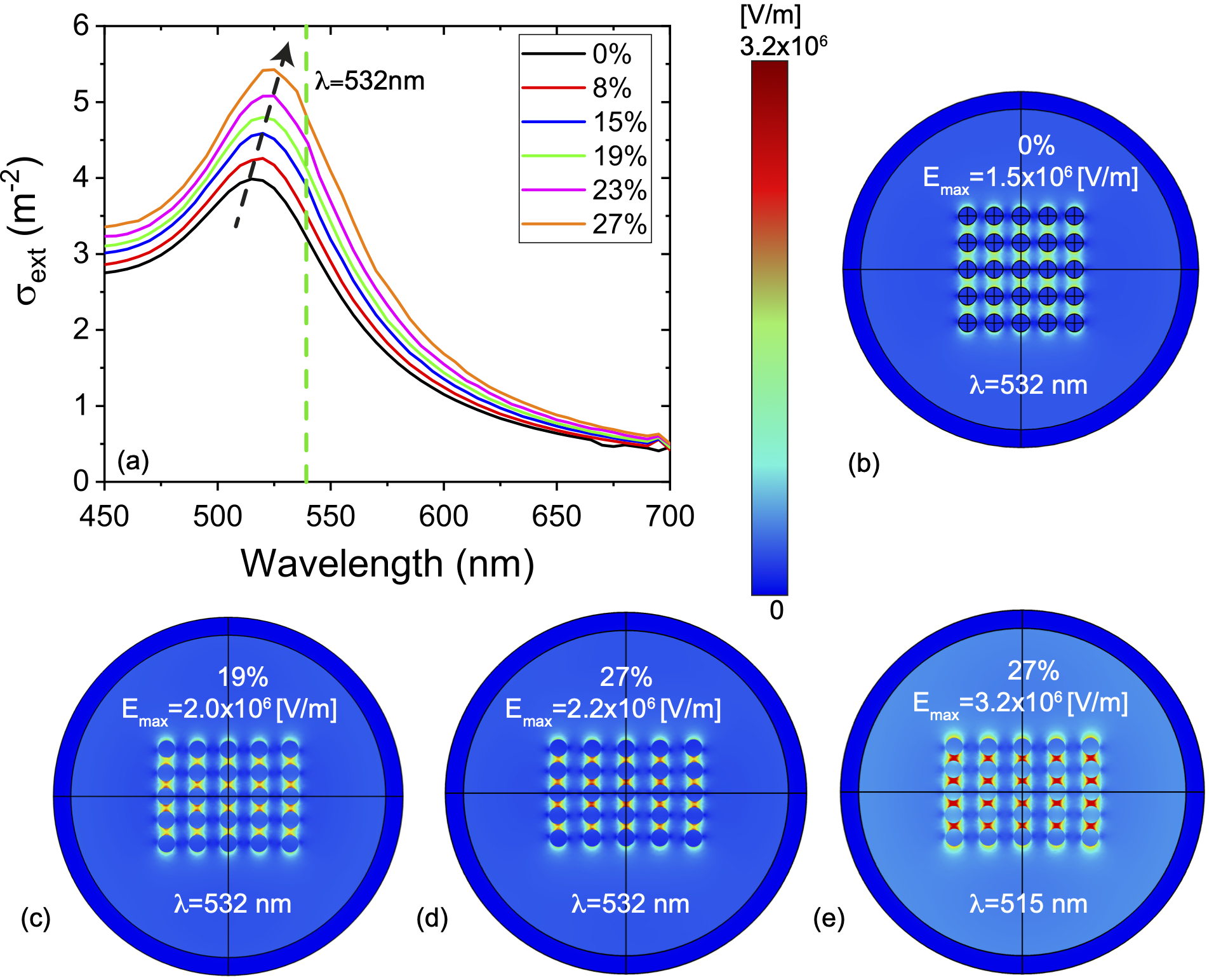}
  \caption{(a) Extinction cross section curves for the different stretching percentages: the rest condition is the black solid line with the plasmonic peak around $\lambda_{p0}$= 510 nm. By stretching the structure the amplitude of $\sigma_{ext}$ increases and the peak shifts until $\lambda_{p27\%}$= 515nm. The green dashed line indicates the wavelength of the used laser in the thermal "numerical experiments" ($\lambda_{therm} = 532 nm$). (b-d) Electric field maps for the rest condition and two different stretching values under the $\lambda_{therm}$ excitation. (e) E maps for the 27\% of stretching at the new $\lambda_{p27\%}$.}
\label{2}
\end{center}
\end{figure}

In Figure \ref{2}a, we report the $\sigma_{ext}$ for different stretching percentages, from the rest condition to the maximum applicable stretching yet avoiding the GNPs overlapping. It is possible to observe that by stretching the substrate from 0\% to 27\% a redshift of the $\lambda_{p0}$  from 510 nm to 
$\lambda_{p27\%}$=515 nm occurs. 

The electric field (E) maps, for the GNPs arrangement at rest (Figure \ref{2}b) and for the different percentages of stretching (Figure \ref{2}c-d), show the different enhancement and distribution of the electric field around the NPs as a function of the particles displacement. These maps are collected at $\lambda$=532 nm, that corresponds to the typical excitation wavelength source, used in laboratory, to excite photo-thermal effects in spherical NPs. 
As we can see, by increasing the stretching percentage we obtain an enhancement of the E field due to the interaction between the nanoparticles that are closer than in the rest condition. 
It is clear that the electromagnetic coupling between the GNPs follows the spectral behavior of the extinction curves. In fact, if we plot the E map for the max stretching percentage considered (27\%) at the new $\lambda_{p27\%}$= 515 nm we obtain an intensity of E$_{max}$ equal to 3.2 $\cdot$ 10$^6$ V/m, that is even higher with respect to the value obtained at 532 nm (E$_{max}$=2.2 $\cdot$10$^6$ V/m).
\subsection{Thermal analysis}
\begin{figure}[h]
\begin{center}
\includegraphics[width=1\columnwidth]{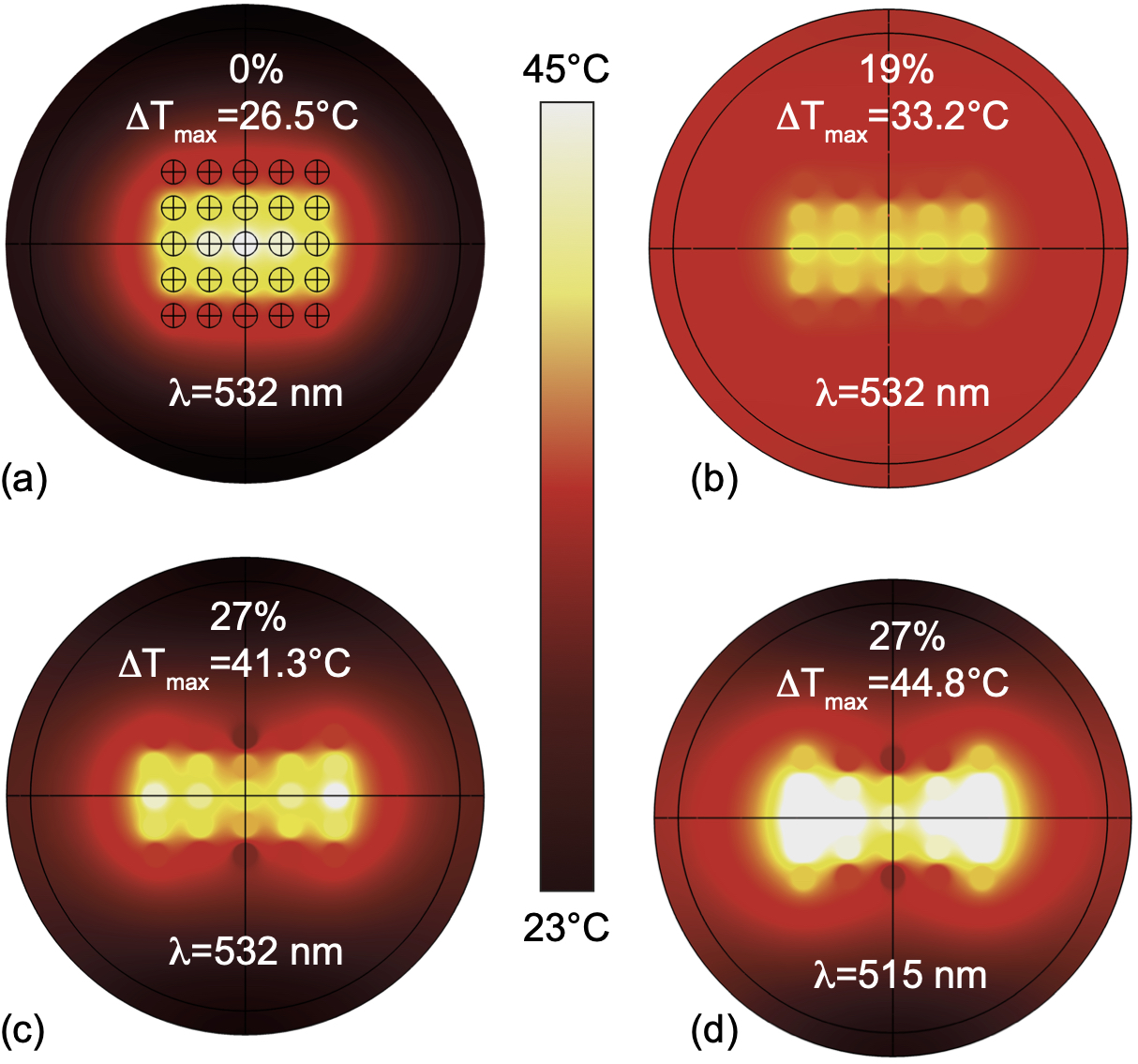}
  \caption{Thermal maps of the (a) rest, (b) 19\%, (c) 27\% stretched structure at $\lambda_{therm}$. (d) Thermal map for the max stretching percentage of 27\% at $\lambda_{p_{27\%}}$.}
\label{3}
\end{center}
\end{figure}

After stretching the structure and collecting the modification of its plasmonic response due to the corresponding change in $\sigma_{ext}$, the last part of the work is devoted to characterize the GNPs arrangement in terms of photo-induced heating. The square structure allows the formation of chains that interact as electromagnetic and thermal hot spots resulting in a temperature enhancement. In Figure \ref{3}a-c,  we reported the thermal heating maps ($\Delta$T) of the GNPs distribution for different stretching percentages calculated at  $\lambda_{therm}$. The temperature variation $\Delta T$ gradually increases following the cross section amplitude: in fact, a $\Delta T_{MAX}$=26.5$^\circ$ for the substrate at rest (0\%) is modified in a much higher $\Delta T_{MAX}$=41.3$^\circ$ in case of a 27\% strain percentage. Noticebably, by calculating the $\Delta T$ increase for the wavelength $\lambda_{p27\%}$ corresponding to the highest $\sigma_{ext}$ value, $\Delta T_{MAX}$ is the highest with (44.8$^\circ$).
\section{Conclusions}
In this contribution, we presented a comprehensive electromagnetic and thermal analysis of 2D square arrangement of plasmonic nano-structures. The proposed numerical simulations are obtained by using COMSOL Multiphysics for the entire experimental process from the stretching to the photo-thermal evaluation.  By using these coupled multi-physics models we pave the route to study and prototype, in a simple and fast way, tunable/active plasmonic systems that can work like nano-heaters. This kind of studies evidence specific regions of thermal hot-spots and thus envision the possibility to realize devices in a wide scenario of applications including medical, environmental and biological field to name a few. 
\section*{Methods}
By the use of the mechanical tool, we study how an applied external strain can deform the initial structure. To use it, we fix the force (\textbf{F}) that represents the boundary load, and the relative constrains. Generally, the constrains involved for a stretching case are the following: the edges along the $y$ direction are left free to move, and they are fixed as zero in $x$ and $z$ directions to avoid bend and shear actions. For the same reasons the edges in the $x$ direction are free to move along $x$ and are constrained in $y$ and $z$ directions. To reproduce arrangements characterized by different displacements, it is necessary to change step by step \textbf{F} and save the corresponding geometry as \textit{.stl} file. The resulting new geometry, that corresponds to the stretched structure, should be imported in the electromagnetic-thermic model to follow the process to study the $\sigma_{ext}$ and $\Delta$T.  
The extinction cross sections and temperature variations, evaluated for different stretching percentages, have been obtained by implementing a 3D simulation in \textit{COMSOL Multiphysics}. The simulation involves a controlled spherical volume, also called perfect matched layer (PML), whose size is much larger than the radius of the NPs, $r_{PML}$ = 20R or 30R, depending on the specific geometry. 

The absorption and scattering cross sections are calculated, in the \textit{Electro-Magnetic Wave, Frequency Domain (emw)}€ module, by means of the following relations: $\sigma_{abs}$=$W_{abs}/P_{in}$, $\sigma_{sca}$=$W_{sca}/P_{in}$, where $P_{in}$ is the incident irradiance, defined as energy flux of the incident wave; $W_{abs}$ is the energy rate absorbed by particle that is derived by integrating the energy loss $Q_{loss}$ over the volume of the particle, while $W_{sca}$ is the energy rate absorbed by particle that  is derived by integrating is derived by integration the Poynting vector over an imaginary sphere around the particle.
All the other parameters as scattered field method, proper polarization and propagation direction as opportunely set to numerically investigate the near field distribution and the plasmonic response of the GNPs.
The light beam intensity,  has been evaluated as $I= E_0^{2}/(2 Z_{0const})$ = 1.33 E7 W/m$^{2}$, where E$_0$ is the initial electric field (in our case 1E5 V/m) and $Z_{0const} $= 376.73 $\Omega$ is the impedance of the system that takes into account also the incident area. This tool package offers the possibility to study the energy flow that passes through the NPs when they are shone by a light beam and the energy flow scattered from NPs and collected on the external layer used as an integrating sphere. 

Finally, the \textit{Heat Transfer in Solids}€ module is used to study the interaction between light and NPs in terms of localized heating. Here it is necessary to set the area that interacts with the electromagnetic wave (NP surfaces and surrounding medium) and the room temperature $T_0$.

The materials involved in these simulations are gold, air and PDMS. For each one, in the \textit{Materials} section of the software the following data have been considered:

\textit{Gold}: real and imaginary part of permittivity as a function of wavelength from the \textit{Materials} library of the software, permeability $\mu$ = 1, the electrical conductivity $\sigma = 1$, Young module $E=7\cdot 10^{10} Pa$, Poisson ratio $\nu=0.44$, thermal conductivity $K=314 W/m\cdot K$, the density $\rho=19300 kg/m^{3}$ and the heat capacity $C_p =126 J/kg\cdot K$.

\textit{Air}: real and imaginary part of permittivity as a function of wavelength, permeability $\mu = 1$, the electrical conductivity $\sigma = 1$, thermal conductivity $K=1 W/m\cdot K$, the density $\rho= 1 kg/m^{3}$ and the heat capacity $C_p =1 J/kg\cdot K$.

\textit{PDMS}: real and imaginary part of permittivity as a function of wavelength, permeability $\mu = 1$, the electrical conductivity $\sigma = 1$, Young module $E=7.5\cdot 10^5 Pa$, Poisson ratio $\nu=0.49$, thermal conductivity $K=0.16 W/m\cdot K$, the density $\rho=970 kg/m^{3}$ and the heat capacity $C_p =1460 J/kg\cdot K$.

\section*{Acknowledgements}
The authors thank the ``Area della Ricerca di Roma 2", Tor Vergata, for the access to the ICT Services (ARToV-CNR) for the use of the COMSOL Multiphysics Platform and Origin Lab, and the Infrastructure ``BeyondNano" (PONa3-00362) of CNR-Nanotec for the access to research instruments.

\section*{References}
\bibliography{bibliograph_gio} 
\end{document}